\documentclass[showpacs, pra,onecolumn,preprintnumbers ,amsmath, amssymb, superscriptaddress, aps]{revtex4-2}
\usepackage{color}
\usepackage{amsmath,amssymb}
\usepackage{pifont}
\usepackage{amssymb}  
\usepackage{bbold}
\usepackage{float}
\usepackage{subfloat}

\usepackage[caption=false]{subfig}
\usepackage{tikz}
\usepackage{makecell}
\usepackage{subfig}
\usepackage{pifont}   
\usepackage{graphicx} 
\graphicspath{{Figures/}}
\usepackage{dcolumn}  
\usepackage{bm}       
\usepackage{multirow} 
\usepackage{placeins}
\usepackage[colorlinks]{hyperref}
\usepackage{mathtools}
\usepackage{appendix}

\begin{document}

\title{Energy levels of  graphene magnetic  quantum dot in inhomogeneous gap}
\date{\today}
\author{Fatima Belokda}
\affiliation{Laboratory of Theoretical Physics, Faculty of Sciences, Choua\"ib Doukkali University, PO Box 20, 24000 El Jadida, Morocco}
\author{Ahmed Jellal}
\email{a.jellal@ucd.ac.ma}
\affiliation{Laboratory of Theoretical Physics, Faculty of Sciences, Choua\"ib Doukkali University, PO Box 20, 24000 El Jadida, Morocco}
\affiliation{Canadian Quantum  Research Center,
	204-3002 32 Ave Vernon,  BC V1T 2L7,  Canada}

\author{El Houssine Atmani}
\affiliation{Laboratory of Nanostructures and Advanced Materials, Mechanics and Thermofluids, Faculty of Sciences and Techniques, Hassan II University, Mohammedia, Morocco}


			\begin{abstract}
		We investigate the energy levels of charge carriers confined in a magnetic quantum dot in graphene with an inhomogeneous gap through an electrical potential. We solve the eigenvalue equation for two regions. We explicitly determine the eigenspinors for both valleys $ K $, $ K' $ and use the boundary condition at the quantum dot interface to obtain the energy levels. We show that the energy levels exhibit symmetric and asymmetric behavior under appropriate conditions of the physical parameters. It has been found that changing the energy levels by introducing an energy gap outside the quantum dot changes the electrical properties. 
			\end{abstract}

	\maketitle

	\section{Introduction}
Graphene is defined as a two-dimensional layer composed only of carbon atoms arranged in a honeycomb pattern \cite{Novoslov1}.
The structure and the nature of the bonds between the atoms that compose it give graphene extraordinary and unique electronic properties \cite{Novoselov2,Castro}.
The dispersion relation is linear such that the valence band and the conduction band touch at two points, K and K', called Dirac points. At these points, the graphene electrons behave as massless Dirac fermions \cite{Castro,Geim A} and are described by the Dirac equation \cite{Castro}. In the framework of band theory, graphene is a zero-gap semiconductor. In general, the ultra-relativistic nature of graphene's charge carriers has led researchers to question how they would respond to confinement \cite{Gomes Peres 4}. It is precisely this particular property that prohibits the use of fabrication techniques. The question of confining Dirac fermions in graphene has led to many proposals. There are several theoretical methods for confining Dirac fermions in single-layer graphene \cite{Rozhkov Giavaras,Chen}, such as inhomogeneous magnetic fields \cite{Martino, Espinosa-Ortega19}, potential cylindrical symmetry \cite{Chen}, spatial modulation of the Dirac gap \cite{Giavaras Franco},  cutting the flake into small nanostructures \cite{Zebrowski Wach,Thomson M R}, using the substrate to induce a band gap \cite{Recher P} and so on.
 
 \par Graphene quantum dots (QDs) are small disk-like pieces of graphene with a radius of $r_0$ in which the electronic wave function is confined and exhibits quantum confinement effects, regardless of size. Therefore, graphene QDs have a non-zero band gap and are luminescent by excitation. This band gap is tunable by changing the size of the graphene QDs. Since its discovery, scientists have been trying to trap electrons in graphene-based QDs in anticipation of the wide field of novel applications of QDs in electronic devices \cite{C. Berger}, valves \cite{Espinosa-Ortega19}, photovoltaics \cite{Bacon M}, qubits \cite{Trauzettel B}, and gas detection \cite{Sun H}. QDs made from nanostructures are very specific to the exact shape of the edge, which is difficult to control \cite{Espinosa-Ortega19}.

\par Motivated by our previous work \cite{Belokda}, we study the influence of two inhomogenous gaps on the energy spectrum of the graphene QDs. For this, we consider the confinement of charge carriers in a magnetic graphene quantum dot with the  gap  $\Delta_{1}$ inside and the gap $\Delta_{2}$ outside the dot. To obtain the eigenspinors, we solve the Dirac equation separately in each region of the system. By using the boundary condition at the interface of the quantum dot, we obtain an equation describing the energy levels as a function of the physical parameters.  We numerically study the energy levels as a function of the radius of QDs, the magnetic field, the electrostatic potential, and two gaps $\Delta_{1}$, $\Delta_{2}$.

\par This paper is organized as follows. In Sec. \ref{MTH}, we set up the mathematical tools needed to study the present system. In Sec. \ref{EEE}, we  determine the eigenspinors that describe the fermions in our theoretical model. Using the boundary condition, we derive a formula governing the energy levels as a function of the physical parameters in Sec. \ref{LLL}.  We numerically analyze the energy levels in Sec. \ref{RDI}. Finally, we conclude our results.

	\section{Model and theory} \label{MTH}
	
\begin{figure}[h]
	\centering
	\subfloat{\includegraphics[scale=0.3]{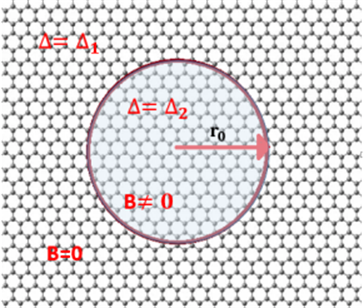}}
\caption{(color online)  A circular quantum dot of radius $r_0$ surrounded by a graphene sheet in the presence of a perpendicular magnetic field $B $ inside and inhomogeneous gaps is depicted schematically. }
	\label{11}
\end{figure}

Let us consider the magnetic confinement of Dirac fermions in graphene subjected to an electrostatic potential and two inhomogeneous gaps.  
We create a quantum dot with radius $r_0$ in the presence of two different gaps, one inside the other
\begin{equation}
	\Delta(r)=\left\{%
	\begin{array}{ll}
		\Delta_{1}, &  r>r_0  \\
		\Delta_{2} , &  r < r_0 
	\end{array}\right.
\end{equation}
using the magnetic field along $z$-direction shown below	
	\begin{equation} \label{BBFF}  
		\vec{B}=
		\left\{%
		\begin{array}{ll}
			B\vec{e}_z, & r<r_0 \\
			0, &  r> r_0. 
		\end{array}
		\right. 
	\end{equation}
Such a system can be illustrated, as shown in  Fig. \ref{11}. 
%

The following Hamiltonian can be used to describe the dynamics of carriers in the honycomb lattice of covalently bound carbon atoms in a single graphene 
	\begin{equation}\label{ham1}
		H = v_{F}(\vec{p}+e\vec{A})\cdot \vec{\sigma}+V(r) \mathbb{I}_2+\Delta{(r)} \sigma_z  
		 \end{equation} 
where the Fermi velocity is $ v_{F} \approx 10^6$ m/s, $\vec p = (p_x, p_y) $ is the two-dimensional momentum operator, $\vec \sigma$ denotes the Pauli matrices, $\mathbb{I}_2$ is the $2$ $\times$ $2$ identity matrix,  and $V$ is the applied potential. The  vector potential can be calculated using \eqref{BBFF} to get
\begin{equation}	
	\vec A=
	\left\{%
	\begin{array}{ll}
		\dfrac{B}{2r} \left(r^{2} - R^{2}\right)\vec{e_z}, & r<r_0 \\
		0 ,&  r > r_0 .
	\end{array}\right.
\end{equation} 
In polar coordinates $ (r,\theta) $, the Hamiltonian \eqref{ham1} reduces to the form 
 \begin{equation} \label{ham2}
 	H =	\begin{pmatrix} V_+ &	\pi^{+} 
	   \\ \pi^{-}
	 & V_{-} 
\end{pmatrix}\end{equation}
and  the momentum operators are introduced 
\begin{equation} 
	\pi^{\pm}=-i\hbar\nu_{F}e^{\pm i\tau\theta}\left(\partial_{r}\pm\dfrac{\tau i}{r}\partial_{\theta}\right)
\end{equation}
where we have defined 
\begin{equation} 
	V_\pm=V\pm\Delta_{1}\pm\Delta_{2}.
\end{equation} 
We can now calculate the energy spectrum using the eigenvalues equation,  $ H \psi^{\tau}(r,\theta)=E \psi^{\tau}(r,\theta)$. Noting that the Hamiltonian commutes with the total angular momentum $ J_{z}=L_{z}+\frac{\hbar}{2}\sigma_{z}$, the radial and angular components of the eigenspinors $\psi^{\tau}(r,\theta)$ can be separated. As a result, we have
\begin{equation}   
\psi^{\tau}{(r,\theta)}=e^{im\theta} 
\begin{pmatrix} \Phi_{A}^{\tau}(r)  \\ie^{-i\tau\theta}\Phi_{B}^{\tau}(r) \label{2}
\end{pmatrix} 
\end{equation}
where the angular-momentum quantum number is $m = 0, \pm1, \pm2, \cdots$. $\Phi_{A}^{\tau} $ and $\Phi_{B}^{\tau}$ are wave functions that represent   sublattices $A$ and $B$, respectively. The parameter $\tau$ has two values, $\pm1$, which distinguishes the two valleys $K$ and $K'$. 

\section{Eigenspinors}\label{EEE}
Using the mathematical tools established above, we can now get the solutions to the energy spectrum. Indeed, the equations for the region 1 ($r>r_0$) are obtained
 \begin{align}
 &	\left(\partial_{\rho}+\frac{m\tau}{\rho}\right)\Phi^{\tau}_{A}(\rho)=-\left((\varepsilon-v)+\delta_{1}\right)\Phi^{\tau}_{B}(\rho) \label{99} 
\\
&\left(\partial_{\rho}-\frac{m-\tau}{\rho}\right)\Phi^{\tau}_{B}(\rho)=\left((\varepsilon-v)-\delta_{1}\right)\Phi^{\tau}_{A}(\rho)  \label{1010}
\end{align}
where the variable change  $\rho=\frac{r}{r_0}$ and
  dimensionless units $\varepsilon=\frac{E}{E_{0}}$, $v=\frac{V}{E_{0}}$, $\delta_{1}=\frac{\Delta_{1}}{E_{0}}$ are considered, such that
 $E_{0}=\frac{\hbar\nu_{F}}{r_0}$.
 By injecting \eqref{99} into \eqref{1010}, we   get a second order differential equation for $\Phi^{\tau}_{A}(\rho)$
 \begin{equation}
 \left(\rho^{2}\partial^{2}_{\rho}+\rho\partial_{\rho}-m^{2}+\alpha^{2}_{1}\rho^{2}\right)\Phi^{\tau}_{A}(\rho)=0
\end{equation} 
showing the  Bessel function of the first kind,  regular at the origin,
as a solution
\begin{equation}
	\Phi_{A}(\rho)=C^{\tau}_{1}J_{m}(\alpha_{1}\rho)
\end{equation} 
 where the parameter $  \alpha_{1} $ is given by
 \begin{equation}
  \alpha_{1}=\sqrt{|(\varepsilon-v)^{2}-\delta_{1}^{2}|} 
\end{equation} 
   and $C^{\tau}_{1}$ is a   normalization constant. The second component of spinor can be derived from \eqref{99} as
 \begin{equation}
\Phi_{B}(\rho)= -iC^{\tau}_{1}e^{-i\tau\theta}J_{m-\tau}(\alpha_{1}\rho).
\end{equation}
Finally in region 1, the eigenspinors take the form
\begin{equation}
	\psi^{\tau}_{1}{(\rho,\theta)}= C^{\tau}_{1}e^{im\theta} 
\begin{pmatrix} J_{m}(\alpha_{1}\rho)  \\-ie^{-i\tau\theta}J_{m-\tau}(\alpha_{1}\rho) \label{22}
\end{pmatrix}.
\end{equation}

Remember that there is a magnetic field $B$ in region 2 ($r<r_0$), which causes the momentum operators $\pi^\pm $ to take the following forms 
\begin{equation} 
	\pi^{\pm}=-i\hbar\nu_{F}e^{\pm i\tau\theta}\left(\partial_{r}\pm\dfrac{\tau i}{r}\partial_{\theta}\mp\frac{\tau eBr}{2\hbar}\right).
\end{equation} 
Using   $H\psi^{\tau}_{2}=E\psi^{\tau}_{2}$ to get 
\begin{align}
	&\left(\partial_{\rho}+\frac{m\tau}{\rho}+\tau \beta\rho\right)\Phi^{\tau}_{A}(\rho)=-\left((\varepsilon-v)+\delta_{2}\right)\Phi^{\tau}_{B}(\rho) \label{1616} 
\\
&
\left(\partial_{\rho}-\frac{m-\tau}{\rho}-\tau \beta\rho\right)\Phi^{\tau}_{B}(\rho)=\left((\varepsilon-v)-\delta_{2}\right)\Phi^{\tau}_{A}(\rho)  \label{1717} 
\end{align} 
where $ \beta=\frac{eBr_0^{2}}{2\hbar}$ and $\delta_{2}={\frac{\delta_{2}r_0}{\hbar\nu_{F}}}$ have been set. After substitution of
 \eqref{1616} into \eqref{1717}, one obtains
\begin{equation}
	\left(\rho^{2}\partial^{2}_{\rho}+\rho\partial_{\rho}-m^{2}-2\beta(m-\tau)\rho^{2}+\alpha^{2}_{2}\rho^{2}-\beta^{2}\rho^{4}\right)\Phi^{\tau}_{A}(\rho)=0 \label{1818}
\end{equation} 
and we have involved
\begin{align}
\alpha_{2}=\sqrt{|(\varepsilon-v)^{2}-\delta_{2}^{2}|}.	
\end{align}
To solve  \eqref{1818}, let us take the first component of spinor as
\begin{equation} \Phi_{A}(\rho)=\rho^{|m|}e^{-\frac{\beta\rho^{2}}{2}}\varphi(\beta\rho^{2})\label{1919}
\end{equation} 
and define a new variable $\eta=\beta\rho^{2}$
to end up with  the confluent hepergeometric ordinary differential equation
 \begin{equation} \left(\eta{\partial^{2}_\eta}+(b-\eta){\partial_\eta}-a\right)\varphi(\eta)=0 
\end{equation}
with the quantities
\begin{align}   
&	
 a=-\frac{(\varepsilon-v)^{2}-\delta^{2}_{2}}{4\beta}+\frac{m-\tau+|m|+1}{2}\\
 & b=1+|m|.    
\end{align}
 As a result, we obtain the solution 
 \begin{equation}
 	\Phi_{A}(\rho)=\rho^{|m|}e^{-\frac{\beta\rho^{2}}{2}}C^{\tau}_{2}
 	M\left(a,b,\beta\rho^{2}\right)
 \end{equation}
where $M(a,b,\beta\rho^{2})$ is the confluent hypergeometric function and $ C^{\tau}_{2}$ is a  normalization constant. Thus, the second component can be extracted from \eqref{1616} as
   \begin{equation} \Phi_{B}(\rho)=i\frac{\rho^{|m|}e^{-\frac{\beta\rho^{2}}{2}} e^{-i\tau\theta}}{(\varepsilon-v)+\delta_{2}}\left[\left(\frac{\tau m}{\rho}+\beta\tau\rho\right)M\left(a,b,\beta\rho^{2}\right)-aM\left(a+1,b+1,\beta\rho^{2}\right)\right]. 
   \end{equation} 
   When we combine everything, we get the eigenspinors in region 2
 \begin{equation}   
 	\psi^{\tau}_{2}(\rho,\theta)= C^{\tau}_{2}\rho^{|m|}e^{-\frac{\beta\rho^{2}}{2}}e^{im\theta}\begin{pmatrix} M\left(a,b,\beta\rho^{2}\right)  \\\frac{ie^{-i\tau\theta}}{(\varepsilon-v)+\delta_{2}}\left[\left(\frac{\tau m}{\rho}+\beta\tau\rho\right)M\left(a,b,\beta\rho^{2}\right)-aM\left(a+1,b+1,\beta\rho^{2}\right)\right]\end{pmatrix}.
 \end{equation}
 
 \section{Energy levels}\label{LLL}
We will proceed by using the boundary condition at the interface, $ r = r_0 $, which is equivalent to $\rho = 1$, because obtaining the energy level explicitly for the current system is difficult.  In fact, the operation 
\begin{align}
	\psi^{\tau}_{1}(1, \theta)=\psi^{\tau}_{2}(1, \theta)
\end{align}
 produces the relations
\begin{align} &C^{\tau}_{1}J_{m}(\alpha_{1})=C^{\tau}_{2}e^-\frac{\beta}{2} M(a,b,\beta) 
\\
& C^{\tau}_{1}J_{m-\tau}(\alpha_{1}) = -C^{\tau}_{2}\frac{e^{-\frac{\beta}{2}}}{(\varepsilon-v)+\delta_{2}}
\left[({\tau m}+\tau\beta)M(a,b,\beta)-aM(a+1,b+1,\beta)\right].
\end{align}
It is convenient to write this set of equations in the matrix from 
\begin{equation}   
	M^{\tau} \begin{pmatrix} C^{\tau}_{1}\\C^{\tau}_{2} \end{pmatrix}=0 \end{equation}
where we have defined
 \begin{equation}   M^{\tau}= 	\begin{pmatrix} J_{m}(\alpha_{1}) &	 e^{-\frac{\beta}{2}} M(a,b,\beta)
			\\ J_{m-\tau}(\alpha_{1})
			& -\frac{e^{-\frac{\beta}{2}}}{(\varepsilon-v)+\delta_{2}}[({\tau m}+\tau\beta)M(a,b,\beta)-aM(a+1,b+1,\beta)]\end{pmatrix}. \end{equation}
	By requiring a null determinant, $ \det M_\tau = 0 $, we can now determine the equation governing the energy level. This process yields
		  \begin{equation} -J_{m}(\alpha_{1})\frac{e^{-\frac{\beta}{2}}}{(\varepsilon-v)+\delta_{2}}[({\tau m}+\tau\beta)M(a,b,\beta)-aM(a+1,b+1,\beta)]-J_{m-\tau}(\alpha_{1})e^{-\frac{\beta}{2}} M(a,b,\beta)=0 \label{2929}.
		 \end{equation}
Since this involves several physical parameters, a numerical analysis is required to underline the basic features of the energy levels. In fact, we will discuss the results shown in various plots of energy under appropriate conditions.

\section{Results and discussion}\label{RDI}

\begin{figure}[H]
\centering
\includegraphics[scale=0.35]{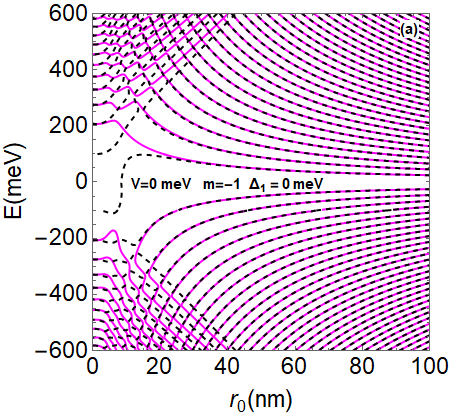}	
\includegraphics[scale=0.35]{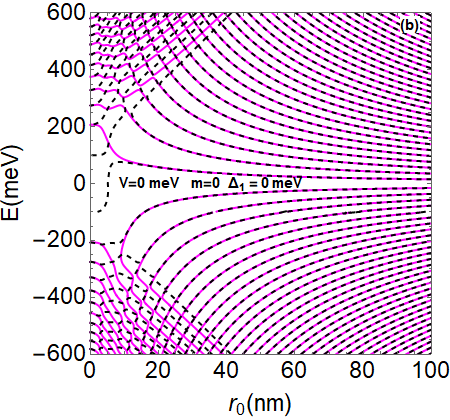}
					\includegraphics[scale=0.35]{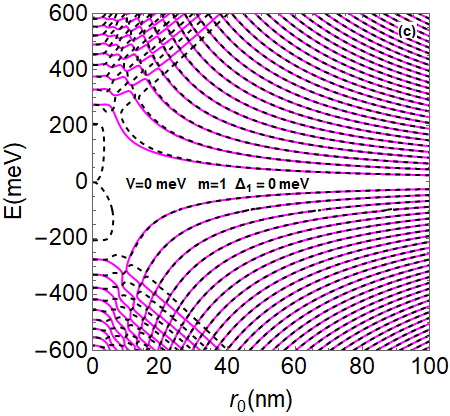}\\
					\includegraphics[scale=0.35]{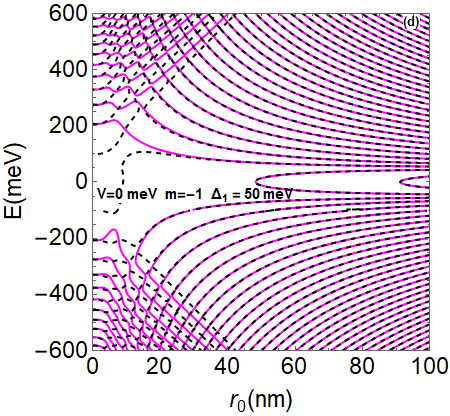}
					\includegraphics[scale=0.35]{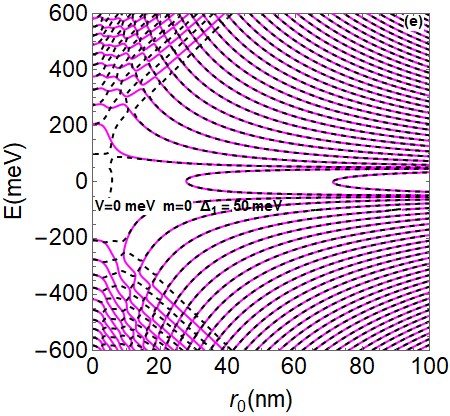}
				    \includegraphics[scale=0.35]{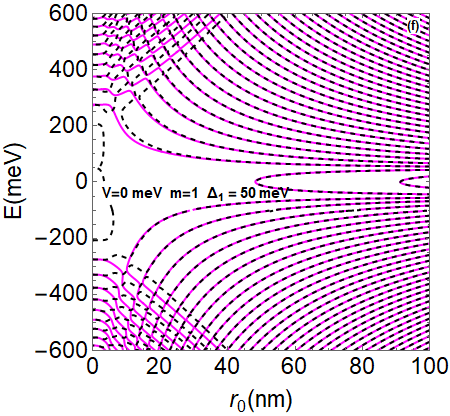}\\
					\subfloat{\includegraphics[scale=0.35]{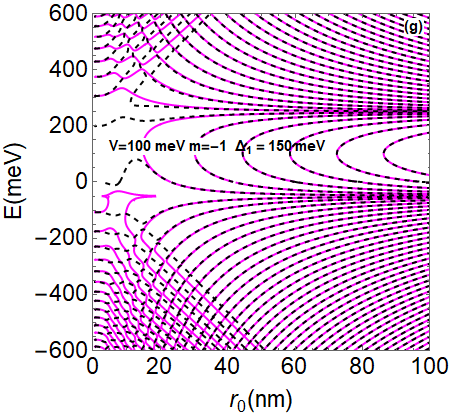}}
					\subfloat{\includegraphics[scale=0.35]{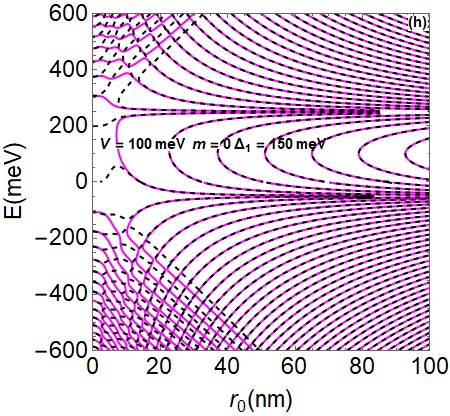}}	\subfloat{\includegraphics[scale=0.35]{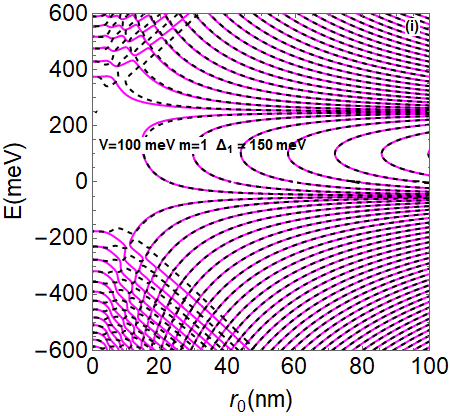}}
				\caption  {(color online) Energy levels as a function of the quantum radius $r_0$ for $B=  25$ T, $\Delta_{2}=100$ meV and $m= -1,0,1$. (a,b,c):  $\Delta_{1}=0$ and V= $0$ meV,  (d,e,f):  $\Delta_{1}=50$ meV   and V= $0$ meV,  and (g,h,i): $\Delta_{1}=150$ and  V= $100$ meV.  The magenta curves:  $\tau=1$ and the dashed black curves: $\tau=-1$. }\label{f2}	
		 	\end{figure}

The energy levels for a quantum dot in graphene as a function of radius $r_0$ for a magnetic field $B = 25 $ T and a gap $\Delta_2 = 100 $ meV inside the quantum dot are shown in Fig. \ref{f2}. We choose three angular momentum $ m $ values (a,d,g): $m = -1 $, (b,e,h): $m=0$, (c,f,i): $m= 1 $, as well as three gap values outside the quantum dot  (a, b,c): $\Delta_1= 0 $ meV, $\Delta_1= 50 $ meV, and $\Delta_1= 150 $ meV. Here, the blue curves represent valley $ K $ ($\tau=1$) and the dashed black curves represent valley $ K' $ ($\tau=-1$). 
We see that for very small values of $ r_0 $, the energy levels for both valleys, $ K $ and $ K' $, are degenerate. They demonstrate the symmetry $E(m, \tau) = - E(m, -\tau)$ as well as the asymmetry $E(m, \tau) \neq E(m, -\tau)$. When the value of $r_0$ approaches $40$ nm, the energy levels become linear and exhibit the symmetry $E(m, \tau) = E(m, -\tau)$.  
Furthermore, for $\Delta_1 = 50$ meV, Figs. \ref{f2}(d,e,f) show a number of new levels between the valence and conductance bands, particularly for higher $r_0$ values. In Figs. \ref{f2}(g,h,i), we see that the number of levels increases and occupies the area of $-\Delta_1+V < E <\Delta_1 +V $. When compared to the results obtained in \cite{ Belouad Jellal,Mirzakhani M}, these show new features due to the presence of the gap $\Delta_1$. Figs. \ref{f2}(g,h,i) show that the energy levels move vertically with respect to the potential $V$, which is consistent with \cite{Belouad Jellal} observations. 

\begin{figure}[H]
	\centering
		\includegraphics[scale=0.35]{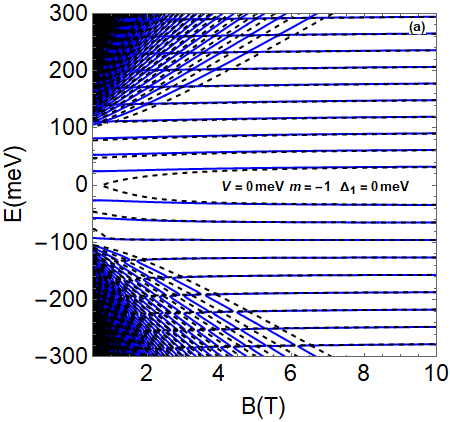}
		\includegraphics[scale=0.35]{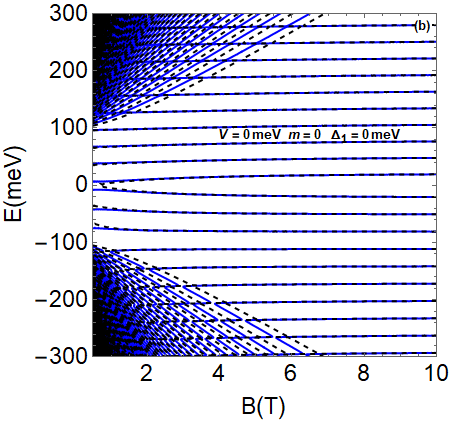}  \includegraphics[scale=0.35]{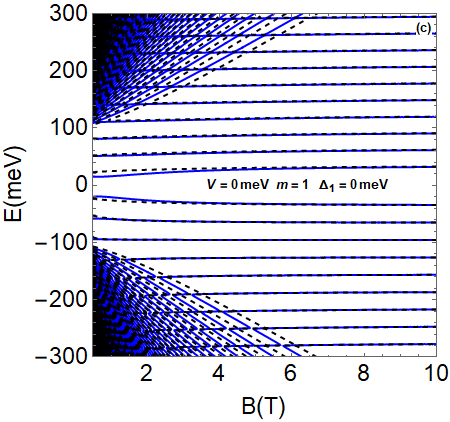}\\
		\includegraphics[scale=0.35]{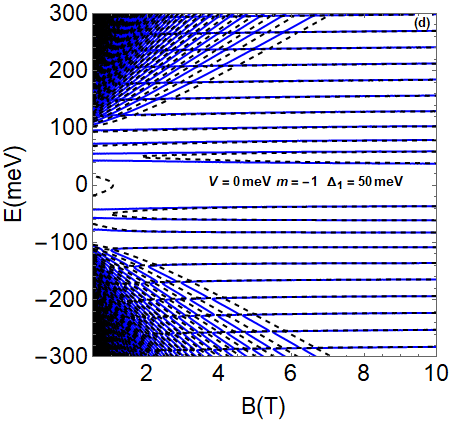}
		\includegraphics[scale=0.35]{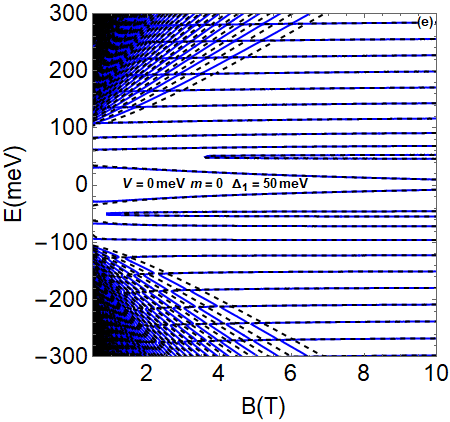} \includegraphics[scale=0.35]{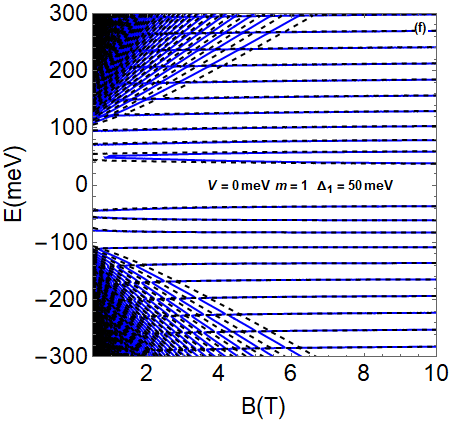}\\
\subfloat{\includegraphics[scale=0.35]{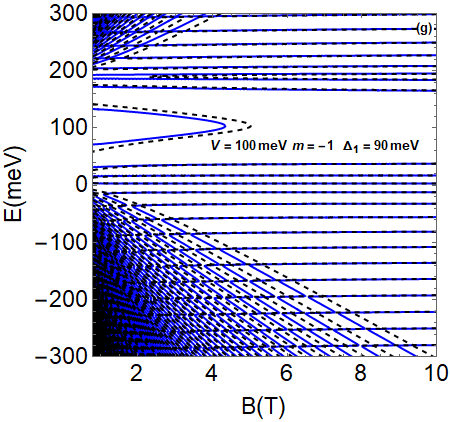}}
\subfloat{\includegraphics[scale=0.35]{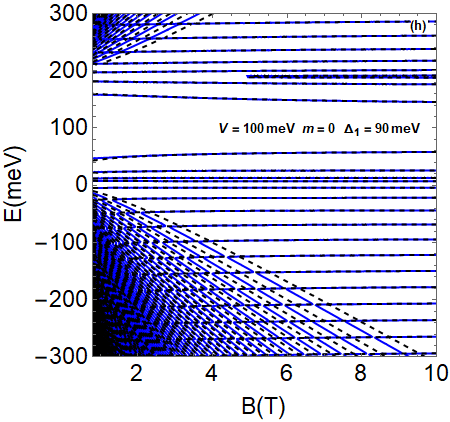}}	\subfloat{\includegraphics[scale=0.35]{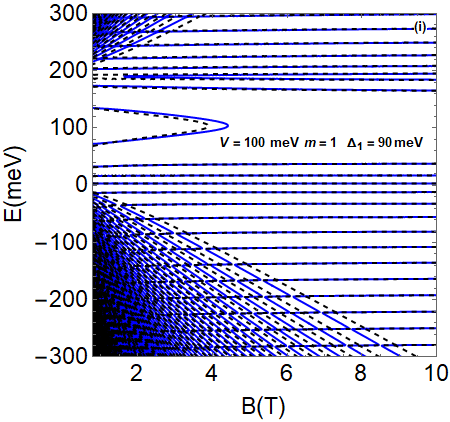}}

\caption  {(color online) Energy levels as a function of the magnetic field $B$ with $r_0=70$ nm  and $m =  -1,0,1$. (a,b,c):  $\Delta_{1}=0$ meV  and $V = 0$ meV,  (d,e,f): $\Delta_{1}=50$ meV  and V = $0$ meV,   (g,h,i): $\Delta_{1}=90$ meV and V = $100$ meV.   Blue curves: $\tau$= $1$ and  dashed black curves: $\tau$=$-1$. }\label{f3}
\end{figure}

Fig. \ref{f3} shows the energy levels as a function of magnetic field $ B $ for $r_0=70 $ nm, $\Delta_2=100 $ meV, and $m = -1,0,1 $, where (a,b,c): $\Delta_1=0 $ meV and $V = 0 $ meV,  (d,e,f): $\Delta_1=50 $ meV and $V = 0 $ meV,  (g,h,i): $\Delta_1=150 $ meV and $V = 100 $ meV. The blue   curves represent valley $ K $ ($\tau=1$), while the dashed black lines represent valley $ K' $ curves ($\tau=-1$).
We find that the energy levels vary according to three regimes, depending on the values of the two gaps, $\Delta_{1}$ and $\Delta_{2}$. 
In  the first case  $\Delta_2<E< - \Delta_2$, when $ B$ is small,  the energy levels become degenerate. By increasing  $B$ we notice that the degeneracy is removed by maintaining  the same symmetry as obtained in \cite{Belouad Jellal}. 
For the second case {$\Delta_1<E<\Delta_2$ or $-\Delta_2<E<-\Delta_1$}, we observe that the energy levels are linear and non-symmetric.
In the third case, $-\Delta_1<E<\Delta_1$, new energy levels appear for very small values of $B$, breaking the symmetry for the angular momentum  $m\neq$ $0$, as shown in Figs. \ref{f3}(g,i). We can see that increasing $\Delta_1$ increases the number of energy levels and fulfill the symmetry $E(m, \tau) = E(-m, -\tau)$. 
These levels disappear as $B$ increases, and for $m = 0$, regardless of $B$ value, there is an energy difference of $2\Delta_1$ between the valence band and the conductance band. 

\begin{figure}[H]
	\centering
	\includegraphics[scale=0.37]{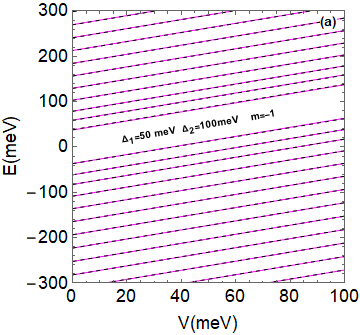}
	\includegraphics[scale=0.37]{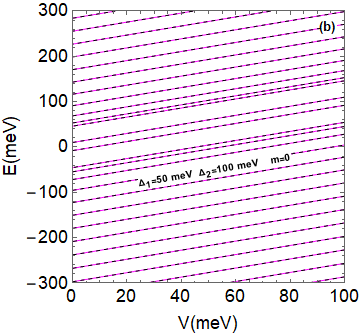}  \includegraphics[scale=0.37]{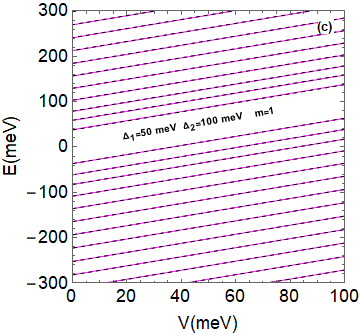}\\
	\includegraphics[scale=0.37]{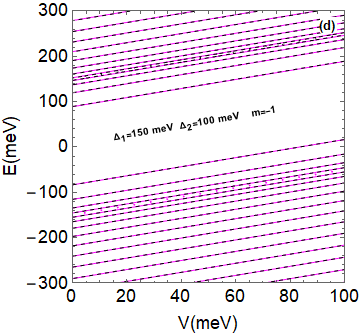}
	\includegraphics[scale=0.37]{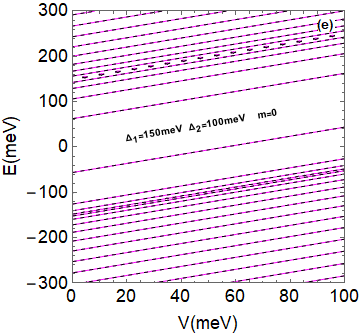} \includegraphics[scale=0.37]{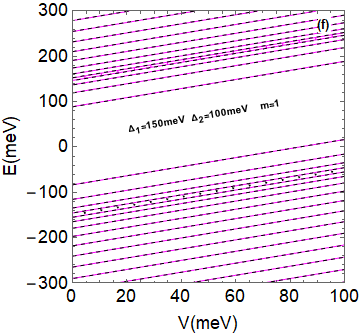}

\caption  {(color online) Energy levels as a function of the potential $ V  $ with $r_0=70$ nm, $\Delta_{2}=100$ meV, $ B =10$ T and  $m= -1,0,1$, with (a,b,c): $\Delta_{1}=50$ meV  and (d,e,f):   $\Delta_{1}=150$ meV.  Magenta curves:  $\tau=1$ and  dashed black curves: $\tau=-1$.  } \label{f4} 
\end{figure}

The energy levels as a function of potential $ V $ are shown in Fig. \ref{f4} for $r_0 = 70$ nm, $\Delta_2 = 100$ meV, $B = 10$ T, and $m = -1,0,{1}$, with (a, b, c): $\Delta_1 = 50$ meV and (d, e, f): $\Delta_1 = 150$ meV. Valley $ K $ ($\tau=1$) is represented by the magenta curves, and valley $ K' $ ($\tau=-1$) is represented by the dashed black curves. We observe that the energy levels are clearly linear and have the symmetry	$E(m,\tau ) =E(m,-\tau)$. 
 There is no energy level for $ m = 0$ in the energy zone $-\Delta_1  <E< \Delta_1$. However, when $E$ is close to zero, there are equidistant levels for $m\neq 0$.
 By increasing $\Delta_{1}$, we see that the gap between the two bands increases. We see a coincidence of energy levels for $E=\pm\Delta_1$. We conclude that the positions of the energy levels are strongly dependent on  the gap $\Delta_{1}$ outside the quantum dot.

\begin{figure}[H]
	\centering

		\includegraphics[scale=0.3]{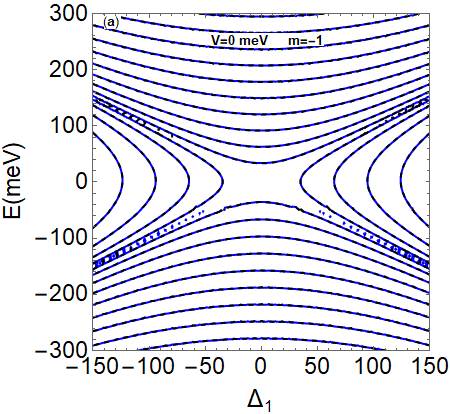}
		\includegraphics[scale=0.3]{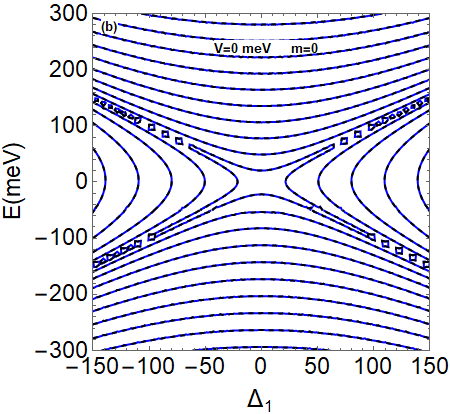}  \includegraphics[scale=0.3]{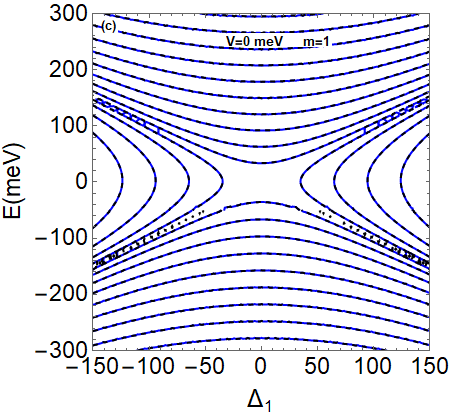}\\
		\includegraphics[scale=0.3]{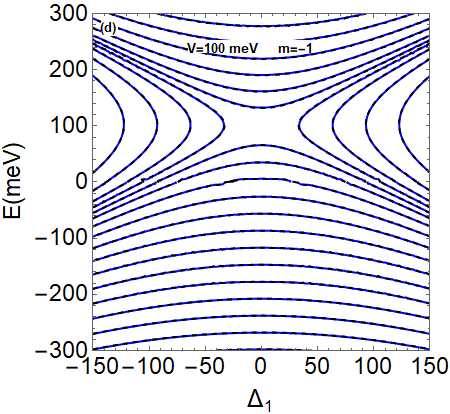}
		\includegraphics[scale=0.3]{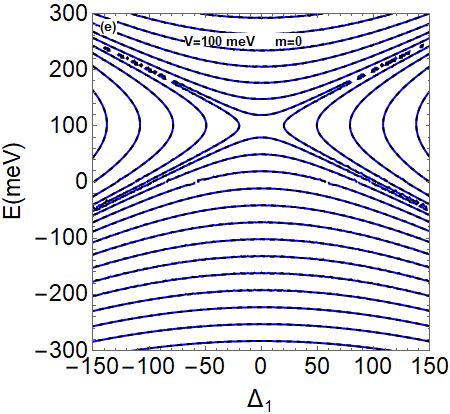} \includegraphics[scale=0.3]{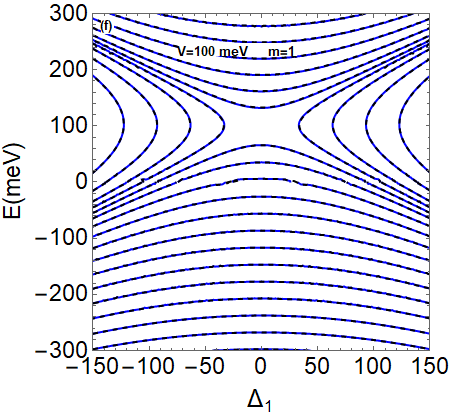}

\caption  {(color online) Energy levels as a function of the gap $\Delta_{1}$ for $B=10$ T, $r_0=70$ nm, $\Delta_{2}=50$ meV  and $ m= -1,0,1$, with  (a,b,c): $V= 0 $ meV  and (d,e,f):  $V=100 $    meV.}\label{f5} 
\end{figure}

The variation of the energy levels as a function of the gap $\Delta_1$ is shown in Fig. \ref{f5} for $B=10$ T, $r_0 = 70 $ nm, $\Delta_2=50 $ meV, and $m = -1,0,1 $, with two potential values (a,b,c): $V = 0 $ meV and (d,e,f): $V = 100 $ meV. The blue curves represent valley $ K $ ($\tau$=1) and the dashed black curves are for  valley $ K' $ ($\tau$=-1). For {-150 meV$>E>150$ meV} (the maximum value of $\Delta_1$),  we observe that
the energy levels vary in horizontal parabolic form and have the symmetry $E(m,\tau) = E(m,\tau)$.
There are new vertical parabolic levels for {-150 meV $<E<150 $ meV}  and 50 meV $<\Delta_{1}< - 50$ meV that differ from those obtained
 in \cite{Farsi A}.


\begin{figure}[H]
	\centering	
		\includegraphics[scale=0.3]{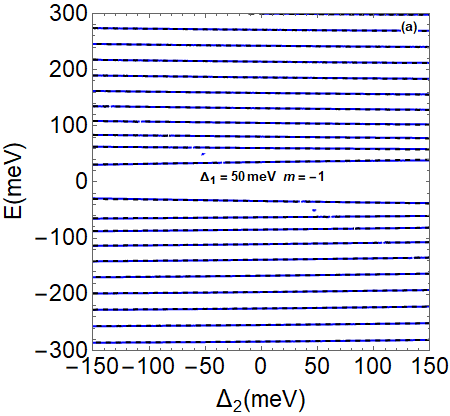}
		\includegraphics[scale=0.3]{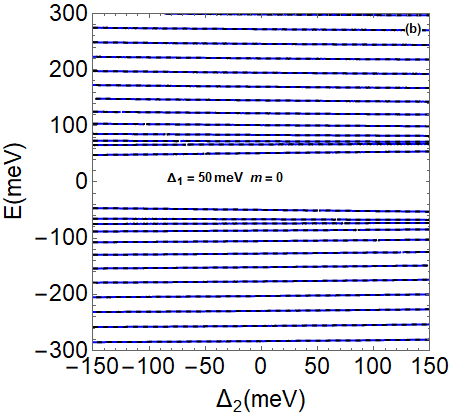}
		\includegraphics[scale=0.3]{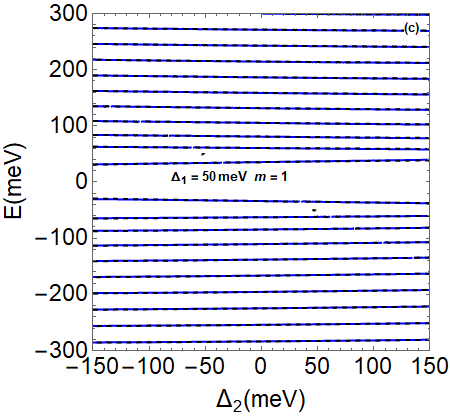}\\
		\includegraphics[scale=0.3]{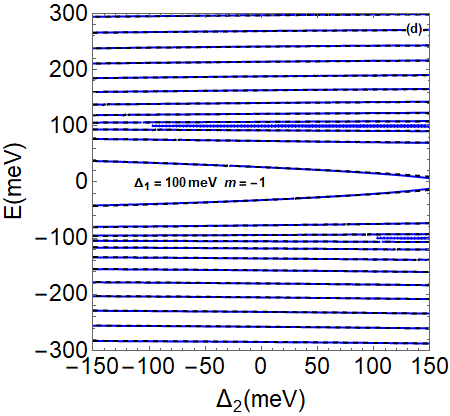}
		\includegraphics[scale=0.3]{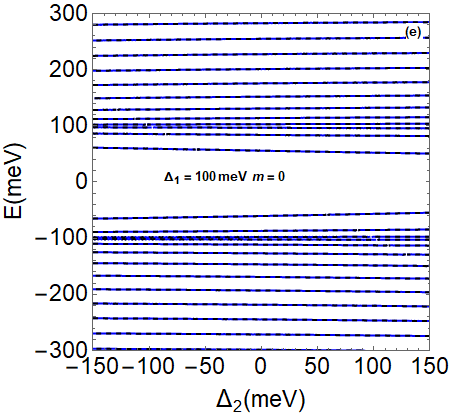} \includegraphics[scale=0.3]{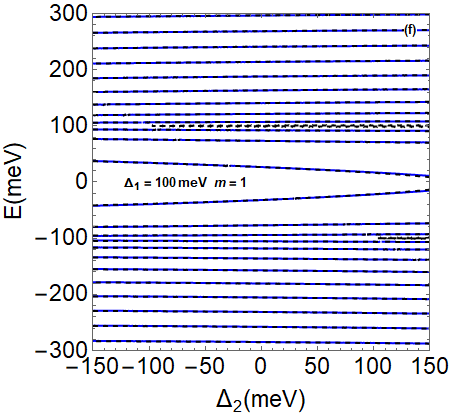}

	\caption  {(color online) Energy levels as a function of the  gap $\Delta_{2}$  for $B=10$ T, $r_0=70$ nm, $V=0$ meV  and $ m= -1,0,1$, with (a,b,c): $\Delta_{1}= 50$ meV and (d,e,f):   $\Delta_{1}=100$  meV.  Blue curves: $\tau=1$ and dashed black curves: $\tau$=-$1$. }\label{f6}  
\end{figure}

In Fig. \ref{f6},  we show the energy levels as a function of the  gap $\Delta_{2}$ for $B=10$ T, $r_0=70$ nm,   $V=0$ meV and  $m =-1,0,1$. with (a,b,c): $\Delta_{1}=50$ meV   and  (d,e,f): $\Delta_{1}$=100 meV. 
Valley $ K $ ($\tau=1$) is represented by the blue curves, and valley $ K' $ ($\tau=-1$) is represented by the dashed black curves. 
We see that the energy levels vary linearly and follow the symmetry $E(m,\tau) = E (m,- \tau)$. 
When $E>\Delta_{1}=50$ meV and $E < - \Delta_{1}=-50$ meV, there is a $2\Delta_1$ gap between the valence and conductance bands for $m=0$, which decreases as $m\neq0$. 
As shown in Fig. \ref{f6}(d,e,f), increasing $\Delta_1$ reduces the gap between two bands for $m=0$, but for $m\neq0$, when $-\Delta_1< E<\Delta_1$, there are energy levels satisfying the symmetry $E(m, \tau) = E (- m, - \tau)$.

%
%
%

\section{Conclusion}

\par In a graphene quantum dot with an inhomogeneous gap in the electrostatic potential and a perpendicular magnetic field, we have investigated the dynamics of charge carriers. We solved the Dirac equation to obtain the eigenspinors inside and outside the quantum dot. We were able to derive an equation describing the energy levels in accordance with the physical parameters that define our system by using the continuity condition at the quantum dot's interface. The obtained energy levels are rich, which pushed us to underline their basic features.

%

\par Our numerical results are shown by studying the energy spectrum. 
Indeed, the study of the spectrum's dependence on the QD's radius $ r_0 $ revealed that the energy spectrum maintained the degeneracy of the asymmetry $E(m,\tau)\neq E(m,-\tau)$ when the limit $r_0 \longrightarrow 0$ is satisfied. 
The energy levels become linear as $ r_0 $ increases, representing the symmetry $E(m,\tau)=E(m,-\tau)$. Furthermore, we see new and different energy levels occupying the region $-\Delta_1<E{<}\Delta_1$, and the number of these levels increases as the gap $\Delta_1$ outside the QD is increased. 
In terms of the energy spectrum's dependence on the magnetic field $B$, we have seen that the spectra for $\pm\Delta_1<E<\pm\Delta_2$ are linear even when $B$ is close to zero. We observed the appearing energy levels as  $\Delta_1$ was increased, with the exception of the angular momentum $m \neq$ 0 and they satisfy the symmetry $E(m, \tau) = E(-m, {-\tau)}$ This means that the transport properties of charge carriers at the center of the graphene magnetic QD are affected by  $\Delta_1$. 
Furthermore, the electrostatic potential $V$ influenced the energy level to shift vertically from $V$. Finally, we have seen that the introduction of  $\Delta_1$ in each spectrum representation has broken the symmetry and resulted in new energy levels of the value energy $-\Delta_1<E<\Delta_1$.

%
%

%

		 \end{document}